\documentclass{sig-alternate-10pt}
\usepackage[normalem]{ulem}
\usepackage{epsfig}
\usepackage{graphicx}
\usepackage{balance}
\usepackage{comment}
\usepackage[hyphens]{url}
\usepackage{cite}
\usepackage{textcomp}
\usepackage{listings}
\usepackage{color,soul}
\usepackage{colortbl}
\usepackage[dvipsnames]{xcolor}
\usepackage{leading}
\usepackage{moreverb}
\usepackage{listings}
\usepackage{framed}
\usepackage{multirow}

  {\begin{list}{$\bullet$}%
     {\setlength{\parsep}{0pt}%
      \setlength{\topsep}{0pt}%
      \setlength{\itemsep}{2pt}}}%
  {\end{list}}

\newcommand\candidate[1]{} 

\newcommand\candidatetwo[1]{} 

\newcommand\itemno[1]{({\em #1})}

\usepackage{subfigure}

\usepackage[english]{babel}
\usepackage{blindtext}

\usepackage{etoolbox}
\makeatletter
\patchcmd{\maketitle}{\@copyrightspace}{}{}{}
\makeatother

\newcommand\copyfromuser{\url{copy_from_user}}
\newcommand\copytouser{\url{copy_to_user}}
\newcommand\mappage{\url{map_page}}
\newcommand\unmappage{\url{unmap_page}}

\begin{document}

\title{\Large \bf Rio: A System Solution for Sharing I/O between Mobile Systems}

\author{
{\large
Technical Report 2013-12-17, Rice University
}
\\
\\
{\large 
     Ardalan Amiri Sani,
     Kevin Boos,
     Min Hong Yun,
     Lin Zhong
}
\\
{\normalsize
     Rice University
}
}

\maketitle

\pagenumbering{arabic}

\section*{Abstract}
\label{sec:abstract}

Mobile systems are equipped with a diverse collection of I/O devices, including cameras, microphones, sensors, and modems. There exist many novel use cases for allowing an application on one mobile system to utilize I/O devices from another. This paper presents Rio, an I/O sharing solution that supports unmodified applications and exposes all the functionality of an I/O device for sharing.  Rio's design is common to many classes of I/O devices, thus significantly reducing the engineering effort to support new I/O devices.  Our implementation of Rio on Android consists of 6700 total lines of code and supports four I/O classes with fewer than 450 class-specific lines of code. Rio also supports I/O sharing between mobile systems of different form factors, including smartphones and tablets. We show that Rio achieves performance close to that of local I/O for audio, sensors, and modems, but suffers noticeable performance degradation for camera due to network throughput limitations between the two systems, which is likely to be alleviated by emerging wireless standards.

\section{Introduction}
\label{sec:introduction}

A user nowadays owns a variety of mobile systems, including smartphones,
tablets, smart glasses, and smart watches, each equipped with a plethora
of I/O devices, such as cameras, speakers, microphones, sensors, and
cellular modems. There are many interesting use cases in which an application running
on one mobile system accesses I/O on another system, for three
fundamental reasons.  \itemno{i} Mobile systems can be in different
physical locations or orientations. For example, one can control
a smartphone's high-resolution camera from a tablet camera application to
more easily capture a self-portrait.
\itemno{ii} Mobile systems can serve different users.  For example, one
can a play music for another user if one's smartphone can access the
other device's speaker.
\itemno{iii} Certain mobile systems have unique I/O devices due to their
distinct form factor and targeted use cases. For example, a user can 
make a phone call from her tablet using the modem and SIM card in 
her smartphone.

Unsurprisingly, solutions exist for sharing various I/O devices, e.g.,
camera~\cite{IP_Webcam}, speaker~\cite{Wi-Fi_Speaker}, modem
(SMS)~\cite{MightyText}, and graphics~\cite{Miracast}.
However, these solutions have three fundamental limitations.  \itemno{i}
{\em They do not support unmodified applications.} For example, IP
Webcam~\cite{IP_Webcam} and MightyText~\cite{MightyText} do not allow
existing applications to use a camera or modem remotely; they only
support their own custom applications.  \itemno{ii} {\em They do not
expose all the functionality of an I/O device for sharing.}  For example, IP
Webcam does not support remote configuration of various camera
parameters, such as depth of focus, exposure, resolution, and white
balance. MightyText supports SMS and MMS from another device, but not
phone calls. Wi-Fi Speaker~\cite{Wi-Fi_Speaker} does not support
configuring equalizer effects on the speaker.  \itemno{iii} {\em They
are I/O class-specific, requiring significant engineering effort to
support new I/O devices.}  For example, IP Webcam~\cite{IP_Webcam} can
share the camera, but not the modem or sensors.

In this paper, we introduce Rio ({\em R}emote {\em I/O}), an I/O sharing
solution for mobile systems that overcomes all three aforementioned
limitations. Rio adopts a split-stack I/O sharing model, in which the
I/O stack, i.e., all software layers from the application to the I/O
device, is split between the two mobile systems at a certain boundary.
All communications that cross this boundary are intercepted on the
mobile system hosting the application and forwarded to the mobile system
with the I/O device, where they are served by the rest of the I/O
stack.  Rio uses {\em device files} as its boundary of choice.  Device
files are used in Unix-like OSes, such as Android and iOS, to abstract many classes of
I/O devices, providing an I/O class-agnostic boundary. The device file
boundary supports I/O sharing for unmodified applications, as it is
transparent to the application layer. It also exposes the full
functionality of each I/O device to other mobile systems by allowing
processes in one system to directly communicate with the device drivers
in another. Rio is not the first system to exploit the device file
boundary; our previous work~\cite{AmiriSani2014} uses device files as
the boundary for I/O virtualization inside a single system.  However,
sharing I/O devices between two physically separate systems engenders
a different set of challenges regarding how to properly exploit this
boundary, as elaborated below.

The design and implementation of Rio must address the following
fundamental challenges that stem from the I/O stack being split across
two systems.  \itemno{i} A process may issue operations on a device file
that require the driver to operate on the process's memory.  With I/O
sharing, however, the process and the driver reside in two different
mobile systems with separate physical memories. In Rio, we support
cross-system memory mapping using a distributed shared memory (DSM)
design that supports access to shared pages by the process, the driver,
and the I/O device (through DMA). We also support cross-system memory
copying with collaboration from both systems.  \itemno{ii} Mobile
systems typically communicate through a wireless connection that has
a high round-trip latency compared to the latency between a process and
driver within the same system.  To address this challenge, we reduce the
number of round trips between the systems due to file operations, memory
operations, or DSM coherence messages.  \itemno{iii} The connection
between mobile systems can break at any time due to mobility or
reliability issues. This can cause undesirable side-effects in the OSes
of all involved systems. We address this problem by properly cleaning up
the residuals of a remote I/O connection upon disconnection, switching
to a local I/O device of the same class, if present, or if not,
returning appropriate error messages to the applications.

We present a prototype implementation of Rio for Android systems. Our
implementation supports four important I/O classes: camera, audio
devices such as speaker and microphone, sensors such as accelerometer,
and cellular modem (for phone calls and SMS). It consists of about 6700 Lines of Code
(LoC), of which less than 450 are specific to I/O classes. It also
supports I/O sharing between heterogeneous mobile systems, including
tablets and smartphones. See~\cite{rio_page} for a video demo of Rio.

We evaluate Rio on Galaxy nexus smartphones and show that it \itemno{i} supports existing
applications, \itemno{ii} allows remote access to all I/O device functionality, 
\itemno{iii} requires low engineering effort to support
different I/O devices, and \itemno{iv} achieves performance close to
that of local I/O for audio devices, sensors, and modem, but suffers
noticeable performance degradation for camera sharing due to Wi-Fi
throughput limitation in our setup and test systems. With emerging wireless standards supporting much higher throughput,
we posit that this degradation is likely to go away in the near future.

\section{Design}

In this section, we describe the design of Rio, including its
architecture and the guarantees it provides to mobile systems using it.

\subsection{Split-Stack Architecture} 

Rio adopts a split-stack model for I/O sharing between mobile systems. 
It intercepts communications at the device file boundary in the
I/O stack on one mobile system and forwards them to the other system to
be executed by the rest of the I/O stack.  

\begin{figure*}[t]
\centering
\subfigure[I/O stack in Unix-like systems]{
\centering
\includegraphics[height=0.15\textheight]{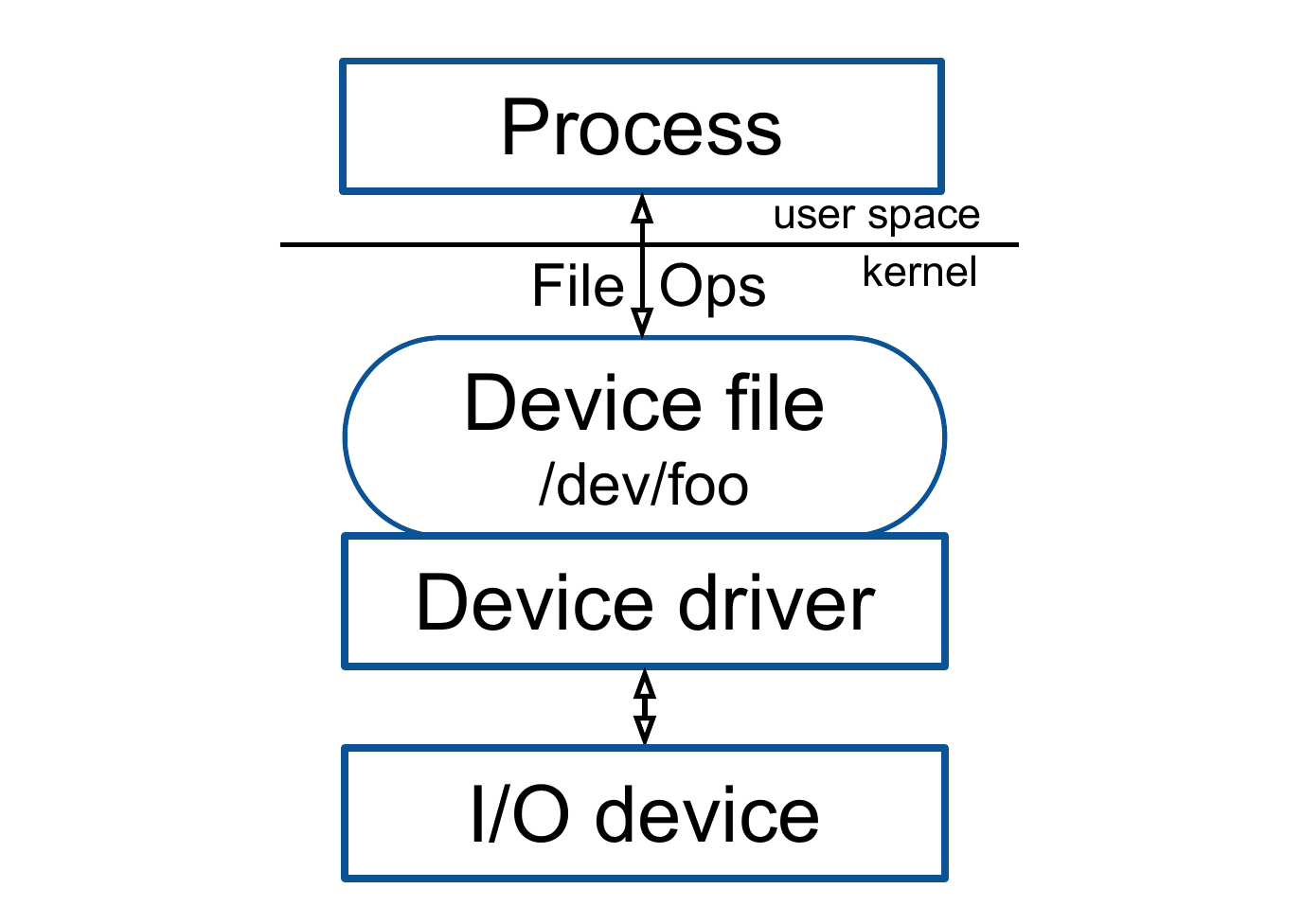}
\label{fig:io_stack}
}%
\hspace{.4in}
\subfigure[Rio splits the I/O stack at the device file boundary]{
\centering
\includegraphics[height=0.15\textheight]{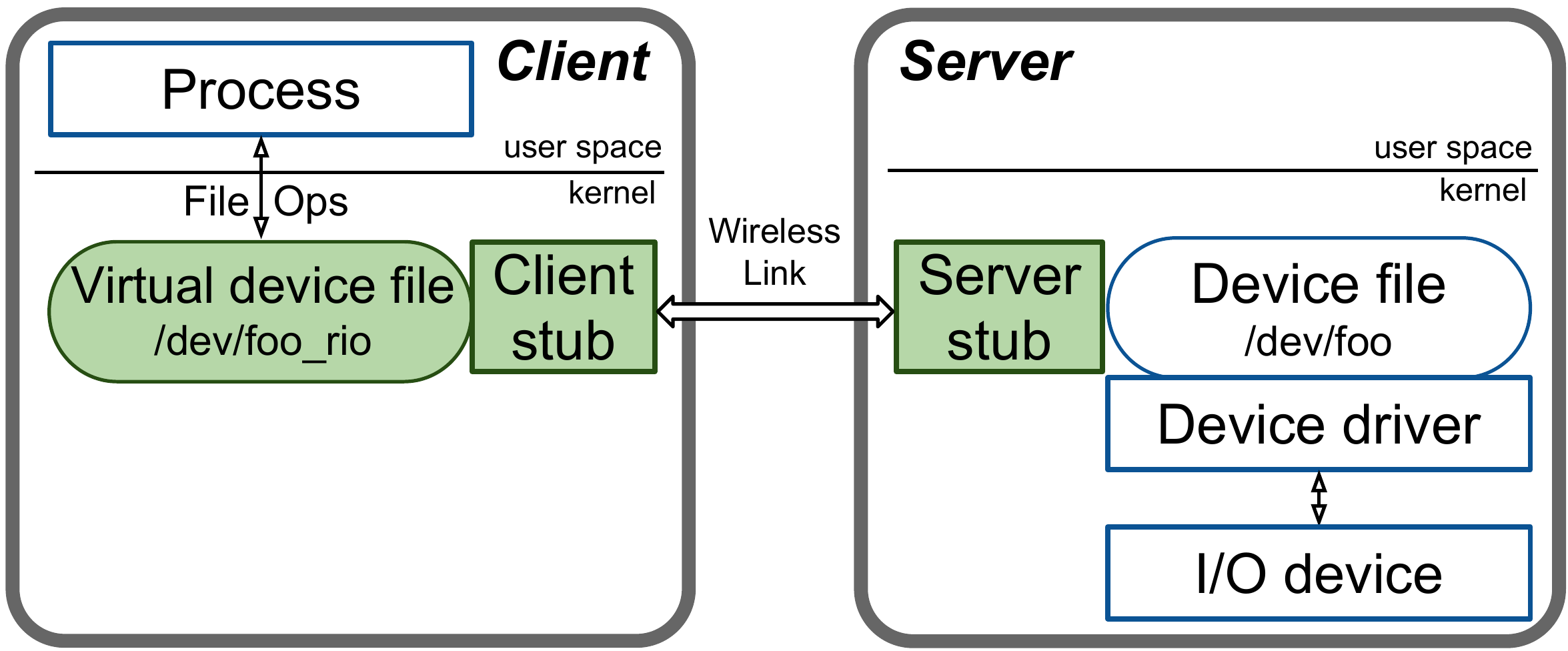}
\label{fig:architecture}
}
\caption{Rio splits the I/O stack at the device file boundary. The process that remotely uses the I/O device
resides in the client system and interacts with a virtual device
file. The actual device file, device driver, and I/O device all reside in
the server system. Rio forwards file operations between the client and server.
The wireless link can either be through an AP or
a device-to-device connection.}
\end{figure*}

Unix-like OSes, such as Android and iOS, use device files to abstract many
classes of I/O devices.  Figure~\ref{fig:io_stack} shows the typical I/O
stack in Unix-like OSes. The device driver runs in the kernel, manages
the device, and exports the I/O device functions to  user space processes
through device files. A process in user space then issues file
operations on the device file in order to interact with the device
driver. The main advantage of using device files as the I/O sharing
boundary is that it is common to many classes of I/O devices, reducing
the engineering effort required to support various I/O classes. 
Moreover, such a boundary is transparent to the application layer and immediately supports
existing applications. It also allows processes to directly
communicate with the driver, hence exposing all I/O device
functionality to other mobile systems.


Figure~\ref{fig:architecture} depicts how Rio splits the I/O stack. It
shows two mobile systems: the {\em server} and the {\em client}. The server
system has an I/O device that the client system wishes to use. 
Rio creates a {\em virtual device file} in the
client that corresponds to the actual device file in the server. The
virtual device file creates the illusion to the client's processes that
the I/O device is present locally on the client. 
To use the remote I/O device, a process in the client executes file operations 
on the virtual device file. These
file operations are intercepted by the {\em client stub} module, which
packs the arguments of each file operation into a packet and forwards it
to the {\em server stub} module. The server stub unpacks the arguments
and executes the file operation on the actual device file.  It then
sends back the return values of the file operation to the client stub,
which returns them to the process.  Note that
Figure~\ref{fig:architecture}  only shows one client using a single I/O
device from a single server.  The design of Rio, however, allows
a client to use multiple I/O devices from multiple servers. It also
allows multiple clients to use an I/O device from a server.

In Rio, the client process is always the initiator of communications
with the server driver.  This is because communications between the
process and driver are always initiated by the process via a file
operation. When an I/O device needs to notify a process of events, the
notification is done using the {\tt poll} file operation.
To wait for an event, a process issues a blocking {\tt
poll} file operation that blocks in the kernel (and hence, in the
server kernel in Rio) until the event occurs. Or, it periodically issues non-blocking
{\tt polls} to check for the occurrence of an event.

Some file operations, such as {\tt read}, {\tt write}, {\tt ioctl}, and
{\tt mmap} require the driver to operate on the process memory. {\tt
mmap} requires the driver to map some memory pages into the process
address space. For this, Rio uses a DSM design that supports access to
shared pages by the client process as well as the server driver and
device (through DMA) (\S\ref{sec:memory_map}). The other three file
operations ask the driver to copy data to or from the process memory.
The server stub intercepts the driver's requests for these copies and
services them with collaboration from the client stub
(\S\ref{sec:memory_copy}).

\subsection{Guarantees}
\label{sec:design_guarantees}

Using I/O remotely at the device file boundary impacts three expected
behaviors of file operations: reliability of connection, latency, and
trust model.  That is, remote I/O introduces the possibility of
disconnection between the process and the driver, adds significant
latency to each file operation due to wireless round trips, and allows
processes and drivers in different trust domains to communicate.
Therefore, Rio provides the following guarantees for the client and
server.

First, to avoid undesirable side-effects in the client and server resulting 
from an unexpected disconnection, Rio triggers a cleanup in 
both systems upon detecting a disconnection.  Rio guarantees
the server that a disconnection behaves similar to  killing a local
process that uses the I/O device. Rio also guarantees the client will 
transparently switch to a local I/O device of the same class, if
possible; otherwise, Rio returns appropriate error messages to the
application (\S\ref{sec:disconnection}).

Second, Rio reduces the number of round trips due to file or memory
operations and DSM coherence messages (\S\ref{sec:latency}) in order to
reduce  latency and improve  performance. Moreover, it guarantees
that the additional latency of file operations only impacts the
performance, {\em not the correctness}, of I/O devices. Rio can provide this
guarantee because most file operation do not have a time-out threshold, but
simply block until the device driver handles them.  {\tt poll} is
the only file operations for which a time-out can be set by the process.
In \S\ref{sec:implementation_specific}, we explain that {\tt poll}
operations used in Android for I/O devices we currently support do
not use the {\tt poll} time-out mechanism. We also explain how Rio can
deal with the {\tt poll} time-out, if used. 

Finally, processes typically trust the device drivers with which they
interact using the device file interface. To maintain that trust, we intend the
current design of Rio for I/O sharing between trusted mobile systems
only.
In \S\ref{sec:discussions_untrusted}, we discuss how the current design
can be enhanced to maintain this guarantee while supporting I/O sharing
between untrusted mobile systems.


 
%
%
%

\section{Cross-System Memory Support}
\label{sec:memory}


In order to handle file operations, the device driver often needs
to operate on the process memory by executing {\em memory operations}.
However,
these operations pose a challenge for Rio because the process and the
driver reside in different mobile systems with separate physical memories.
In this section, we present our solutions.

There are three types of memory operations. The first
one is \mappage, which the driver uses to map
system or device memory pages into the process address space. This memory
operation is used for handling the {\tt mmap} file operation and its
supporting {\tt page\_fault}. Note that the kernel itself
performs the corresponding \unmappage\ memory operation and not the
driver.
The other two types of memory operations are
\copytouser\ and \copyfromuser\footnote{We refer to these operations
using the names of existing Linux functions.}, which the
driver uses to copy a buffer from the kernel to the process memory and
vice-versa. These two memory operations are typically used for handling
the {\tt read}, {\tt write}, and {\tt ioctl} file operations.

\subsection{Cross-System Memory Map}
\label{sec:memory_map}



Cross-system memory map in Rio supports the \mappage\ memory operation
across two mobile systems 
using Distributed Shared Memory
(DSM)~\cite{li88, delp88, lenoski90, Carter91, zhou92, schoinas94}
between them. At the core of Rio's DSM 
is a simple write-invalidate protocol, similar to~\cite{zhou92}. The
novelty of the DSM in Rio is that it can support access to the distributed
shared memory pages not only by a process, but also by kernel
code, such as the driver, and also by the device (through DMA).

\begin{figure}[t]
\centering
\includegraphics[width=1\columnwidth]{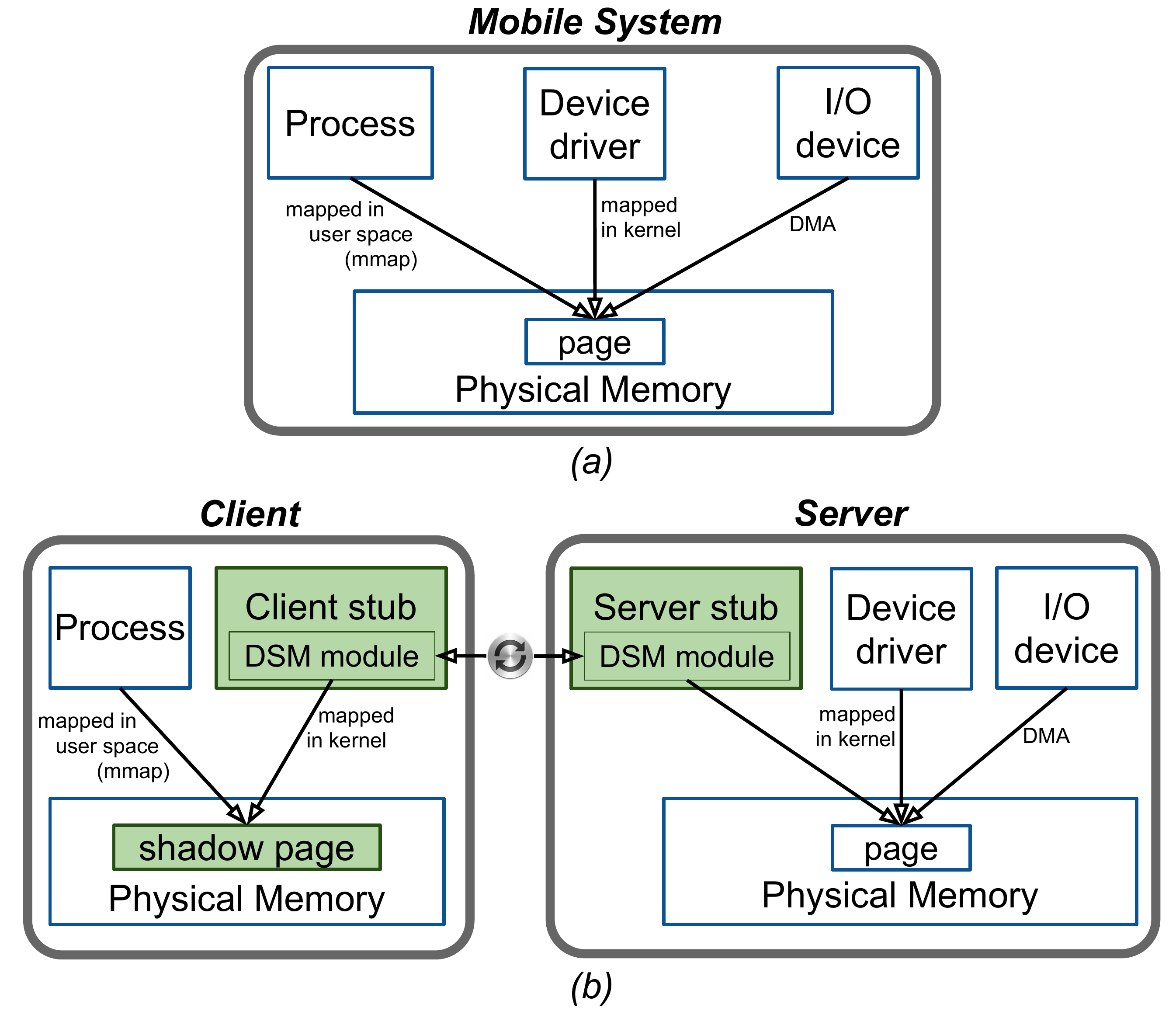}
\caption{(a) Memory map for local I/O device. (b) Cross-system memory
map in Rio.}
\label{fig:map}
\end{figure}

Figure~\ref{fig:map} illustrates the cross-system memory mapping in Rio.
When intercepting
a \mappage\ operation from the server driver, the server stub notifies
the client stub, which then creates a shadow memory page
in the client (corresponding to the actual memory page in the server) and
maps that shadow page into the client process address space. The
DSM modules in these two stubs
guarantee that the process, the driver, and the device have consistent
views of these pages. That is, updates to both the actual and shadow
pages are consistently available to the other mobile system.

We choose a write-invalidate protocol in Rio's DSM for efficiency.
Compared to update protocols that proactively propagate the updates to
other systems~\cite{Carter91}, invalidate protocols do so only when
the updated data is needed on the other system. This minimizes the
amount of data transmitted between the client and server, and therefore
minimizes the resource consumption, e.g., energy, in both systems.
With the invalidate protocol, each memory page can be in one of three possible states:
read-write, read-only, or invalid. 
Although the invalidate protocol is the default in Rio, we can also use an 
update protocol if it offers performance benefits; 
\S\ref{sec:implementation_dsm} explains one such
scenario.

We use 4KB pages (small pages) as the coherence unit because it
is the unit of the \mappage\ memory operation, meaning the driver can map
memory as small as a single small page into the process address space.
When many pages are updated together, 
we batch them altogether to improve
performance (\S\ref{sec:latency_dsm}).

To manage a client process's access to the (shadow) page, we use the page
table permission bits, similar to some existing DSM
solutions~\cite{li88}.  When the shadow page is in the read-write state,
the page table grants the process full access permissions to the page, and
all of the process's read and write instructions execute natively with no
extra overhead. In the read-only state, only write to these pages cause page
faults, while both read and write cause page faults in the invalid
state. Upon a page fault, the client stub triggers appropriate coherence
messages. For a read fault, it fetches the page from the server
and retries the process's operation. For a write fault, it
first fetches the page if in invalid state, and then sends an
invalidation message to the server.

To manage the server driver's access to the page, we use the page table permission bits 
for kernel memory since the driver operates in the kernel.
However, unlike process memory that uses small 4KB pages, certain regions of kernel memory, e.g., the
identity-mapped region in Linux, use larger pages, e.g., 1MB pages
in ARM~\cite{ARM_TRM}, for better TLB and
memory efficiency. 
When the driver requests to map a portion of a large kernel
page into the process address space, the server stub dynamically breaks the large kernel page into
multiple small ones by destroying the old page table entries and
creating new ones. 
With this technique, the server stub can enforce different
protection modes against kernel access to each small page, 
rather than enforcing protection at the granularity of large pages. 
To minimize the side-effects of using small pages in the kernel, e.g., higher TLB
contention, the server stub immediately
stitches the small pages back into a single large page when the pages are unmapped
by the process. 


To manage the server I/O device's access to the page through DMA, the server
stub maintains an explicit state
variable for each page, intercepts the driver's DMA request to the device,
updates the state variable accordingly, and triggers appropriate coherence messages.
Note that it is not possible
to use page table permission bits for a device's access to the page since
devices' DMA operations bypass the page tables. 


Rio supports sequential consistency for two reasons. First, the DSM module
triggers coherence messages immediately upon page faults and DMA
completion, and maintains the order of these messages in
each system. Second, processes and drivers use file
operations to coordinate their own access to mapped pages.

\subsection{Cross-System Copy}
\label{sec:memory_copy}

\begin{figure}[t]
\centering
\includegraphics[width=0.9\columnwidth]{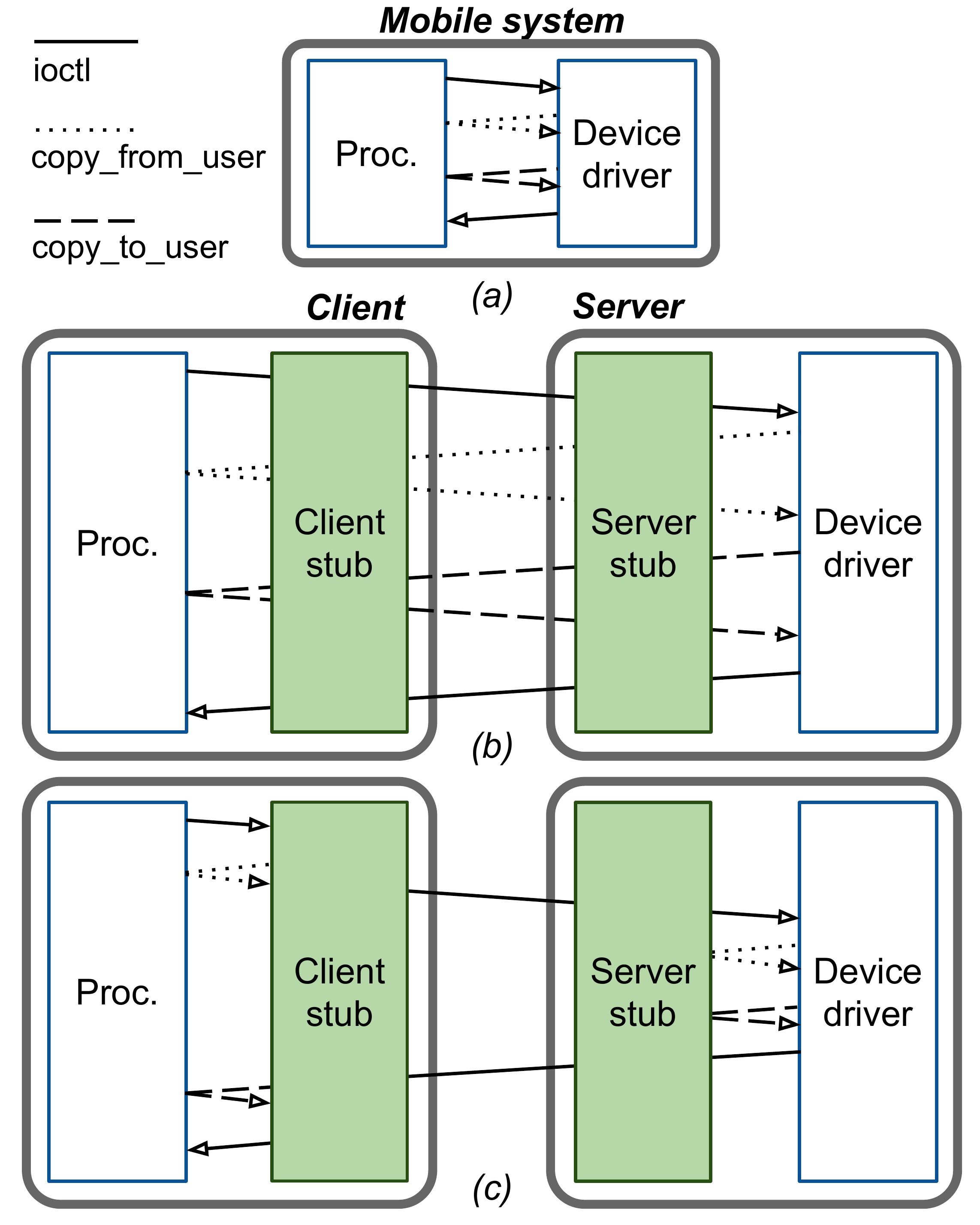}
\caption{Typical execution of an {\tt ioctl} file operation for (a) a local I/O
device, (b) a remote I/O device with unoptimized Rio
(\S\ref{sec:memory_copy}), and (c) a remote I/O device with optimized
Rio (\S\ref{sec:latency_copy}). As the figure shows, the optimization
has reduced the number of round trips from 3 to 1. The reduction can be
more significant in case the file operation requires more memory operations.}
\label{fig:copy}
\end{figure}

Cross-system memory copy in Rio supports the \copyfromuser\ and
\copytouser\ memory operations between two mobile systems.
We achieve this through collaboration between the server and client
stubs.
When intercepting a \copyfromuser\ or \copytouser\ operation from the driver,
the server stub sends a request back to the client stub to perform the
operation.
In the case of \copyfromuser, the client stub copies the data from the
process memory and sends it back to the server stub, which copies it 
into the kernel buffer determined by the driver. In the case of
\copytouser, the server stub copies and sends the data from the kernel buffer to
the client stub, which then copies it to the corresponding process
memory buffer. Figure~\ref{fig:copy} (b) illustrates the
cross-system copy for a typical {\tt ioctl} file operation.



When handling a single file operation, the driver may execute
several memory copy operations, thus causing as many round trips
between the mobile systems because the server stub has to send a separate
request per copy. Large number of round trips can degrade the
I/O performance significantly.
In \S\ref{sec:latency_copy},
we explain how we reduce these round trips to only a single one per file
operation by prefetching
the copy data in the client stub for \copyfromuser\ operations, as well as
batching the data of \copytouser\ operations in the server stub.


\section{Mitigating High Latency}
\label{sec:latency}

The connection between the client and the server typically has high
latency. For example, Wi-Fi and Bluetooth have about 1-2 ms round-trip
latency at best~\cite{Woodings2006}, which is significantly higher than the few
nanoseconds of latency typical of native communications between
a process and device driver (e.g., system calls).
In this section, we discuss the challenges resulting from such high
latency and present our solutions to reduce its effect on I/O
performance by reducing the number of round trips due to copy memory
operations, file operations, and DSM coherence messages.


%

\subsection{Round trips due to copies}
\label{sec:latency_copy}

Round trips due to \copyfromuser\ and \copytouser\ memory operations
present a serious problem to Rio's performance since a single file
operation may execute several copy memory operations in succession. 
For example, a single {\tt ioctl} in Linux's PCM audio driver may
execute four \copyfromuser\ operations.
To solve
this problem, we use the following two techniques. 
\itemno{i} In the client stub, we determine and prefetch all the data needed by the server driver 
and transmit it together with the file operation. With this technique, all
\copyfromuser\ requests from the driver are serviced locally inside the
server. \itemno{ii} In the server stub, we buffer all data that the
driver intends to copy to the process memory and transmit it to the
client along with the return values of the file operation. With this
technique, all \copytouser\ operations can be executed locally in the
client. Figure~\ref{fig:copy} (c) illustrates these techniques. 

Prefetching the data for driver \copyfromuser\ requests requires the
client stub module to determine {\em in advance} the addresses and sizes of the
process memory data buffers needed by the driver.  This is
trivial for the {\tt read} and {\tt write} file operations, as this
information is embedded in their input arguments.
However, doing so for {\tt ioctl} is non-trivial as the {\tt ioctl}
input arguments are not always descriptive enough.  As mentioned
in~\cite{AmiriSani2014}, many well-written drivers embed
information about some simple driver memory operations in one of the
{\tt ioctl} input arguments, i.e., the {\tt ioctl} command number. In
such cases, we parse the command number in the client stub to infer the
memory operations.  There are cases, however, that the command number
does not contain all necessary information. For these cases, we use a static
analysis tool from our previous work that analyses the driver's source
code to extract a small part of the driver code, which can then be executed
either offline or at runtime in the client stub to infer the parameters of driver
memory operations. \cite{AmiriSani2014} provides more details on our
static analysis tool.



To maintain a consistent view of the process memory for the driver, it
is important to update the prefetched data in the server stub upon
buffering the \copytouser\ data if the memory locations overlap.  The
following code segment shows a relevant example from the Linux PCM audio driver.
The driver first updates a data structure field in the process
memory (using the {\tt put\_user()} function, essentially a \copytouser\ memory operation) 
before copying the whole data structure from  the process memory to the kernel).
Updating the prefetched data when
handling  \copytouser\ ensures that the \copyfromuser\ data does not come form
the stale prefetched data and hence guarantees consistency.


\begin{verbatimtab}

struct snd_xferi xferi;
struct snd_xferi __user *_xferi = arg;
...	
if (put_user(0, &_xferi->result))
	return -EFAULT;
if (copy_from_user(&xferi, _xferi,
			sizeof(xferi)))
	return -EFAULT;

\end{verbatimtab}


\subsection{Round trips due to file operations}

File operations are executed synchronously by each process thread, and
therefore, each file operation needs one round trip.
To optimize performance, the process should
issue the minimum number of file operations possible. Changing the
number of file operations is not always possible or may require
substantial changes to the process source code, e.g., the I/O service
code in Android, which is against Rio's goal of reducing engineering effort. 
However, minimal changes to the process code can occasionally result in noticeable
reductions in file operation issuance,  justifying the engineering
effort. 
\S\ref{sec:implementation_specific} explains one example for
Android audio devices.



\subsection{Round trips due to DSM coherence}
\label{sec:latency_dsm}

As mentioned in \S\ref{sec:memory_map}, we use 4 KB small pages as the DSM
coherence unit in Rio. However, when there are updates to several
pages at once, such a relatively small coherence unit causes several round
trips for all data to be fetched. In such cases, transmitting all updated pages 
together in a single round trip is much more efficient. 
\S\ref{sec:implementation_dsm} explains one example for Android camera.

\subsection{Dealing with {\tt poll} time-outs}

{\tt poll} is the only file operations for which the issuing process can
set a time-out. Since Rio adds noticeable latency to each file
operation, it can break the semantics of {\tt poll} if a relatively
small time-out threshold is used.  So far in our Android implementation,
all I/O classes we support do not use {\tt poll} time-out (i.e.,
the process either blocks indefinitely until the event is ready or uses
non-blocking {\tt poll}s).  If {\tt poll} is used with a time-out, the
time-out value should to be adjusted for remote I/O devices. This can be
done inside the handler for {\tt poll}-related syscalls, such as {\tt
select}. Using the heartbeat round trip time
(\S\ref{sec:disconnection}), the client stub can provide a best estimate of the additional
latency that the syscall handler needs to add to the requested time-out
value.
Processes
typically rely on the kernel to enforce the requested {\tt poll}
time-out; therefore, this approach guarantees that the process function
will not break in the face of high latency. In rare cases that the
process uses an external timer to validate its requested time-out, the
process must be modified to accommodate additional latency for remote
I/O devices.


\section{Handling Disconnections}
\label{sec:disconnection}

The connection between the client and the server may be lost 
at any time due to mobility. If not handled properly, the
disconnection can cause the following problems: render the driver
unusable, block the client process indefinitely, or leak resources,
e.g., memory, in
the client and server OSes. When faced
with a disconnection, the server and client stubs take appropriate
actions,
as described below.

We use a time-out mechanism to
detect a disconnection. At regular intervals, the client stub
transmits heartbeat messages to the server stub, which immediately
transmits back an acknowledgement. If the client stub does not
receive the acknowledgement before a certain threshold, or the server
does not hear from the client,  they both trigger
a disconnection event. 
We do not use the in-flight file operations as
a heartbeat because file operations can take unpredictable amounts of time
to complete in the driver. Determining the best heartbeat interval and time-out
thresholds to achieve an acceptable trade-off between overhead and
detection accuracy is part of future work.

{\bf Handling a disconnection in the server:} 
From the perspective of the server, network 
disconnection is equivalent to killing a local process
that is communicating with the driver. Therefore, just as the OS
cleans up the residuals of a killed process, the server stub cleans up the
residuals of the client process after a disconnection.
For each {\tt mmaped} area and each open file descriptor, the server stub invokes the driver's {\tt
close\_map} handler and {\tt release} file operation handler
respectively, in order for the driver to release the allocated
resources. Finally, it closes the file descriptors and releases its
own bookkeeping data structures as well.

{\bf Handling a disconnection in the client:} 
We take two actions in the client upon disconnection. 
First, we clean up the residuals of the disconnected remote I/O
in the client stub,
similar to the cleanup process in the server. Next, we try to make the
disconnection as transparent to the application as possible. If the
client possesses a local I/O device of the same class, 
we transparently switch to that local I/O device after the disconnection. 
If no comparable I/O device is present, we return appropriate
error messages supported by the API. These actions require
class-specific developments, and \S\ref{sec:implementation_specific}
explains how we achieve this for sensors. Switching to local I/O is
possible for three of the I/O classes we currently support, including
camera, audio, and sensors such as accelerometer. For the modem, the
disconnection means that a phone call will be dropped or not initiated, 
or that an SMS will not be sent; all are behaviors understandable by
existing applications.

\section{Android Implementation}
\label{sec:implementation}



We have implemented Rio for Android OS and ARM architecture. The implementation currently
supports four classes of I/O devices: sensors (including accelerometer), audio devices (including microphone and speaker), camera, and cellular modem (for phone calls and SMS). It consists of about 6700 LoC, fewer than 450 of which are
I/O class-specific as shown in Table~\ref{tab:code}. We have tested the implementation on Galaxy Nexus Android
smartphone running Android 4.2.2 with Linux kernel 3.0, and on Samsung
Galaxy Tab 10.1 tablet running Android 4.2.2 with Linux kernel 3.1. The implementation
can share I/O between systems of different form factors: we have demonstrated this
for sharing sensors between a smartphone and a tablet.

\begin{table}[t!]
\begin{centering}
{\scriptsize
\begin{tabular}{|c|c|l|c|}
\hline
 \bf{Type} & \bf{Total} & \bf{Component} & \bf{LoC} \\
           & \bf{LoC}   &                &          \\ \hline
\multirow{5}{*}{Generic}
&\multirow{5}{*}{6291}
& Server stub   & 2801  \\
&& Client stub   & 1651 \\
&& Shared between stubs   & 647 \\
&& DSM     & 1192 \\      
&& Supporting Linux kernel code & 327 \\ \hline
\multirow{6}{*}{\parbox{0.3in}{Class-specific}}
&\multirow{6}{*}{431}
& Camera:        & \\
&&-    HAL        & 36 \\
&&-    DMA        & 134 \\
&& Audio device & 64 \\
&& Sensor       & 128 \\
&& Cellular modem        & 69 \\ \hline

\end{tabular}
}
\caption{Rio code breakdown.}
\label{tab:code}
\end{centering}
\end{table}

\begin{figure}[t]
\centering
\includegraphics[width=0.9\columnwidth]{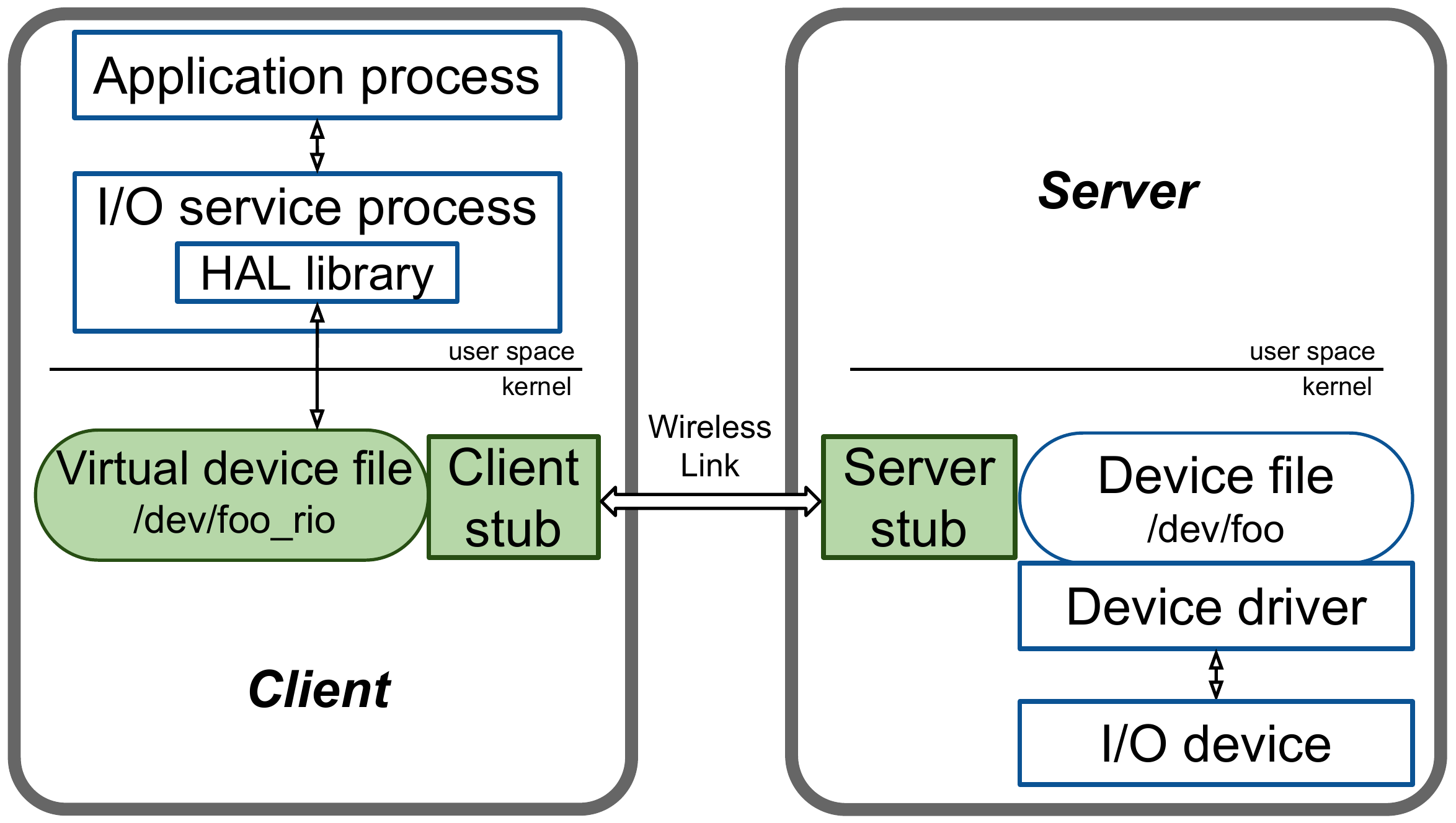}
\caption{Rio's architecture inside an Android system. Rio forwards to the server the file
operations issued by the I/O service process through HAL.
Rio supports unmodified applications but requires small changes to the
class-specific I/O service process and/or HAL.}
\label{fig:android_io_stack}
\end{figure}

Figure~\ref{fig:android_io_stack} shows the architecture of Rio inside an Android system. In Android, the application processes do
not directly use the device files to interact with the driver. Instead,
they communicate to a class-specific {\em I/O service process} through
class-specific APIs. The I/O service process loads a {\em Hardware
Abstraction Layer (HAL)} library in order to use the device file to
interact with the device driver. Rio's device file boundary lies below
the I/O service processes, forwarding its file operations to the server.
As we will explain in the rest of this section, we need small
modifications to the HAL or I/O service process, but no modifications is
needed to the applications.

\subsection{Client \& Server Stubs}

The client and server stubs are the two main components of Rio and
constitute a large potion of Rio's implementation. Each stub has three
modules. The first module supports interactions with applications and device drivers. 
In the client stub, this module intercepts
the file operations and packs their arguments into a  data structure;
in the server stub, it unpacks the arguments from the data structure and
invokes the file operations of the device driver. 
The second module implements the communication with the other stub by serializing data
structures into packets and transmitting them to the other end.
Finally, the third module implements Rio's DSM, further explained in \S\ref{sec:implementation_dsm}.


We use in-kernel TCP sockets for communication between the client and server
stubs~\cite{ksocket}. We use TCP to ensure that all the packets are successfully
received, otherwise the device, driver, or the application might break. 


To handle cross-system memory operations, the server stub intercepts the driver's
kernel function calls for memory operations and forwards them to the
client stub. This includes intercepting 7 kernel functions for \copytouser\ and
\copyfromuser\ and 3 kernel functions for \mappage.
Intercepting the kernel functions supports memory operations from {\em
unmodified drivers}.

\subsection{DSM Module}
\label{sec:implementation_dsm}

Rio's DSM module is shared between the client and the server. It
implements the logic of the DSM protocol, e.g., triggering and handling coherence messages.
The DSM module is invoked in two cases: page faults and DMA. 
We instrument the kernel fault handler to
invoke the DSM module when there is a page fault.
Additionally, the DSM module must handle device DMA to DSM-protected pages.
We monitor the driver's DMA requests to the
device and invoke the DSM module upon DMA completion.

To monitor the driver's DMA requests to devices, we instrument the
corresponding kernel functions. These functions are typically I/O bus-specific,
e.g., I$^2$C, and will apply to all I/O devices using that I/O
bus.
Specialized instrumentation is needed if the driver uses non-standard
interfaces. For example, the camera on the TI OMAP4 SoC inside Galaxy
Nexus smartphones uses custom messages between the driver and the
Imaging Subsystem (ISS) component, where the camera hardware
resides~\cite{OMAP_TRM}. We instrumented
the driver responsible for communications with the ISS to monitor the DMA
requests, only with 134 LoC.



When we receive a DMA completion notification for a memory buffer,
we may use a DSM update protocol to immediately push the
updated buffers to the client an optional optimization. 
Moreover, we update the whole buffer in one round trip. These
optimizations can improve 
performance as they minimize the number of round trips between mobile
systems (\S\ref{sec:latency_dsm}), and hence we used them for camera
frames.

As described in \S\ref{sec:memory_map}, certain regions of the kernel's
address space, namely the identity-mapped region, use large 1MB
pages known as {\em Sections} in the ARM architecture. To split these
1MB Sections into smaller 4KB pages for use with our DSM module, we first walk
the existing page tables to obtain a reference to the Section's
first-level descriptor (a PGD entry). We then allocate a single new page
that holds 512 second-level page table entries (PTEs), one for each
page; altogether, these 512 PTEs reference two 1MB Sections of virtual
memory. We populate each second-level PTE with the correct page frame number and
permission bits from the original Section. Finally, we change the
first-level descriptor entry to point to our new table of second-level
PTEs and flush the corresponding cache and TLB entries.

\subsubsection{Support for Buffer Sharing using Android ION}

Android uses the ION memory management framework to allocate and share
memory buffers for multimedia applications, such as those using the GPU,
camera, and audio~\cite{ION}. The sharing of ION buffers creates unique
challenges for Rio, as demonstrated in the following example.

The camera HAL allocates
buffers using ION and passes the ION buffer handles to the kernel driver,
which translates them to the physical addresses of these buffers and asks the
camera to DMA new frames to them. Once the frames are written, the HAL is
notified and forwards the  ION buffer handle to the graphics
framework for rendering. Now, imagine using a remote camera in Rio. The
same ION buffer handles used by the camera HAL in the client need to be
used by  both the server kernel driver and  the client
graphics framework, since the camera frames from the server are rendered
on the client display.

To solve this problem, we provide support for global ION buffers
that can be used both inside the client and the server. We achieve this
by allocating an ION buffer in the server with similar properties (e.g.,
size) to the
one allocated in the client; we use the DSM module to keep
the two buffers coherent.



\subsection{Class-Specific Developments}
\label{sec:implementation_specific}



Most of Rio's implementation is I/O class-agnostic; our current
implementation only requires under 450 class-specific LoC.




{\bf Resolving naming conflicts}:~~In case the client has an I/O device of the same class that uses device files
with similar names as those used in the server,
the virtual device file must assume a different name
(e.g., {\tt /dev/foo} vs. {\tt /dev/foo\_rio} in
Figure~\ref{fig:architecture}).
However, the device file names are typically hard-coded in the HAL,
necessitating small modifications to use a renamed virtual device file for
remote I/O. 

{\bf Optimizing performance}:~~As discussed in
\S\ref{sec:latency}, sometimes small changes to the HAL
can boost the remote I/O performance significantly by reducing the
number of file operations.
For example, the audio HAL exchanges buffered audio segments with
the driver using {\tt ioctl}s. The HAL
determines the size of the audio segment per {\tt ioctl}.
For local devices (with very low latency), these buffered segments
contain about
3 ms of audio each, less that a round trip time in Rio. 
Therefore, we modify the HAL to use larger buffering segments for remote audio devices.
Although this increases the audio latency, it improves the audio
rate for remote devices. \S\ref{sec:eval} provides measurements to quantify
this trade-off. This modification only required about 30 LoC.

{\bf Support for hot-plugging and disconnection}:~~Remote I/O devices can come and go at any time; in this sense, they behave similarly
to hot-plugging/removal of local I/O devices.  Small changes to the
I/O service layer may be required to support hot-plugging and
disconnection of remote I/O devices.
For example, the Android sensor service
layer opens the sensor device files (through the HAL library) in the phone
initialization process and only uses these file descriptors to
read the sensor values. To support
hot-plugging a set of remote sensors, we modified the sensor service layer to open
the virtual device files and use their file descriptors too when 
remote sensors are present. Upon disconnection, we switch back to using
local sensors to provide an application-transparent mechanism. 

{\bf Avoiding duplicate I/O initialization}:~~Some HAL libraries, including sensor and cellular modem's, perform
initialization of the I/O device upon system boot. However, since the I/O
device is already initialized in the server, the client HAL should not
attempt to initialize the I/O device. Not only this can break the I/O
device, it can also break the HAL because the server device driver rejects
initialization-based file operations. Therefore, the HAL must be
modified in order to avoid initializing a device twice. Achieving this was trivial
for the sensor HAL: we only needed to comment out one LoC. 
However, since the
modem's HAL is not open-source, we had to employ a workaround that uses a second SIM card in the client to initialize the its modem's HAL.
We are developing
a small extension to the modem kernel device driver (which is
open-source) in order to fake the
presence of the SIM card and allow the client HAL to initialize without a second SIM card.

\subsection{Sharing between heterogeneous systems}
Because the device file boundary is common for all Android systems, Rio's design
readily supports sharing between heterogeneous systems, e.g., between a smartphone and a tablet.
However, the implementation has to properly deal with the HAL library of a shared I/O because it may be
specific to the I/O device or to the SoC used in the system. Our solution is to port the HAL library used in
the server to the client. Such a port is easy for two reasons. 
First, Android's interface to the HAL for each I/O class is the same across
Android systems of different form factors. Second, all Android systems use the Linux kernel
and are mostly shipped with ARM processors. For example, we managed to
port the Galaxy Nexus smartphone sensors HAL library to the Samsung Galaxy Tab 10.1
tablet by compiling the smartphone HAL inside the tablet source tree.

\section{Use Cases}


In this section, we
present two sets of use cases of Rio: the first set leverages unmodified
applications with a remote I/O device instead of a local one,
and are immediately demonstrable with Rio (See~\cite{rio_page} for a video demo); the second set requires 
new or modified applications and may need upgrades to the OS I/O stack
components,
e.g., the Android I/O service process (\S\ref{sec:implementation}).

\subsection{Use Cases Demonstrated with Rio}
\label{sec:use_cases_demonstrated}


{\bf Multi-system photography}:~~With Rio, one can use a camera
application on one mobile system to take a photo through a camera on another
system. This capability can be handy in many different scenarios,
especially when taking self-portraits, as it decouples the camera hardware
from the camera viewfinder, capture button, and settings.
Several existing applications try to assist the
user in taking self-portraits using voice recognition, audio guidance, or face
detection~\cite{self_photos}. However, Rio has the advantage in that 
the user can \itemno{i} see the camera viewfinder up close, \itemno{ii} comfortably configure the camera settings, and \itemno{iii} press the capture button whenever
ready, even if the physical camera is dozens of feet away. Alternatively, one can use the front camera on smartphones and
tablets to capture self-portraits, but front cameras capture photos at noticeably lower resolutions compared with rear-facing cameras.

{\bf Multi-system gaming}:~~Larger
mobile systems, such as tablets, provide a superior screen for gaming, but
are bulky compared to pocket-sized smartphones. In addition,
games that use sensors as input, e.g., a racing game, require tilting
the mobile system, making it harder to concentrate on the content of the display. However, with
Rio, a second mobile system, e.g., the smartphone, can be used as a mobile game controller while the larger tablet screen remains stationary.


{\bf One SIM card, many systems}:~~Despite many efforts~\cite{SIM_lock}, users are still
tied to a single SIM card for making and receiving phone calls or SMS,
mainly because the SIM card is associated with a unique number.
With Rio's I/O sharing, the user can make and
receive phone calls and SMS from any of her mobile systems using the
modem and SIM card in her smartphone. For example,
if a user forgets her smartphone at home, she can still receive phone
calls on her tablet at work.

%

{\bf Music sharing}:~~A user might want to allow a friend to listen to some
music via a music subscription application on her smartphone. 
With Rio,
the user can simply play the music on her friend's
smartphone speaker with any existing music playing application.

{\bf Multi-system video conferencing}:~~
When a user is video conferencing on her tablet, she can use the speaker
or microphone on her smartphone by moving them closer to her mouth for
better audio quality in a noisy environment. In a related scenario, she can use the
camera of the smart glasses as an external camera for the tablet to provide a different viewpoint. 




\subsection{Future Use Cases of Rio}
With Rio, new applications can be developed to use the I/O devices available
on another system. 

{\bf Multi-user gaming}:~~The multi-system gaming use case explained in
the previous subsection combined with modifications to the application
can enable novel forms of multi-user gaming across  mobile systems. 
For example, two players can use their smartphones to wirelessly control
a racing game on a single tablet in front of them.
The smartphones' displays can even show in-game context menus or game
controller keys, similar to those on game consoles, providing
a familiar and traditional gaming experience for users.

{\bf Music sharing}:~~
If supported by the audio service process and HAL, a user
can play the same music on her and her friend's smartphones
simultaneously.  With proper application support, the user can even play
two different sound tracks on these two systems at the same time.

{\bf Multi-system video conferencing}:~~A video conferencing application can be extended to show
side-by-side video streams from the smart glasses and the tablet simultaneously. 
In this fashion, the user can not only
share a video stream of her face with her friend, but she can also
share a stream of the scenes in front her at the same time.

{\bf Multi-camera photography}:~~
By using
cameras available on multiple mobile systems, one can realize various 
computational photography techniques~\cite{Raskar2006}. For example, a user can employ the
cameras of her smartphone and smart glasses simultaneously to capture
photos with different exposure times in order to remove motion
blur~\cite{Yuan2007}, or to double the temporal/spatial resolution of video by
interleaving/merging frames from both cameras~\cite{Wilburn2005,
Park2003}.
One can even use the smartphone as an external flash for the smart
glasses camera.

\section{Evaluation}
\label{sec:eval}

We evaluate Rio and show that it \itemno{i} supports legacy
applications, \itemno{ii} allows access to all I/O device functionality,
\itemno{iii} requires low engineering effort to support different I/O
devices, and \itemno{iv} achieves performance close to that of local I/O
for audio devices, sensors, and modem, but exhibits performance drops
for the camera due to throughput limitations. We further discuss that
future wireless standards will eliminate this performance issue.

\begin{figure}[t]
\centering
\includegraphics[width=0.98\columnwidth]{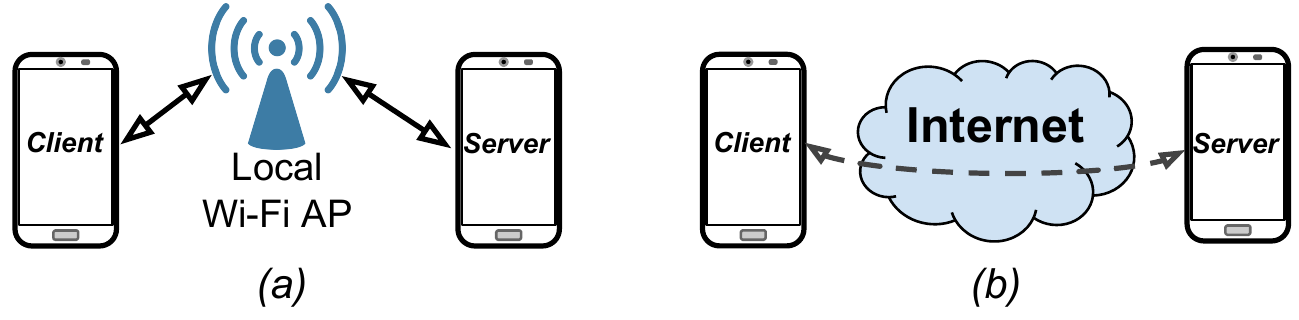}
\vspace{-2mm}
\caption{Evaluation setup.}
\vspace{-2mm}
\label{fig:evaluation_setup}
\end{figure}

All experiments are performed on two Galaxy Nexus smartphones. We use
two connections of differing latency between the phones for the
experiments. The first connection (Figure~\ref{fig:evaluation_setup}(a))
is over wireless LAN between mobile systems that are close to each other,
e.g., both carried by a user or in the same room. We connect both phones
to the same Wi-Fi access point with a median and an average latency of
4.4 ms and 13.8 ms and
14.3 Mbps throughput. 
The second connection (Figure~\ref{fig:evaluation_setup}b) is between
mobile systems at different geographical locations, one at home and one
20 miles away at work. We connect these two phones through the Internet
using external IPs from commodity Internet Providers.  This connection
has a median and average latency of 55.2 ms and 56.9 ms and 1.2 Mbps throughput. 
All reported results use the first LAN connection, unless otherwise
stated.


\subsection{Non-performance Properties}

First, Rio supports existing unmodified applications. We have tested Rio with various
default and third-party applications using different classes of I/O
devices.


Second, unlike existing solutions, Rio exposes all
functionality of remote I/O devices. For example, the client system can
configure every camera parameter, including resolution, exposure,
focus, and white balance. Similarly, an application can configure the speaker with different
equalizer effects.

Supporting new I/O devices in Rio requires low engineering effort. As
shown in Table~\ref{tab:code}, we only needed 128, 64, 170, and 69 LoC
to support sensors, audio devices (both speaker and
microphone), camera, and the modem (for phone calls and SMS), respectively.

\subsection{Performance Benchmarks}

\begin{figure*}[th]
\centering
\vspace{-2mm}
\subfigure[ ]{
  \centering
  \includegraphics[width=0.9\columnwidth]{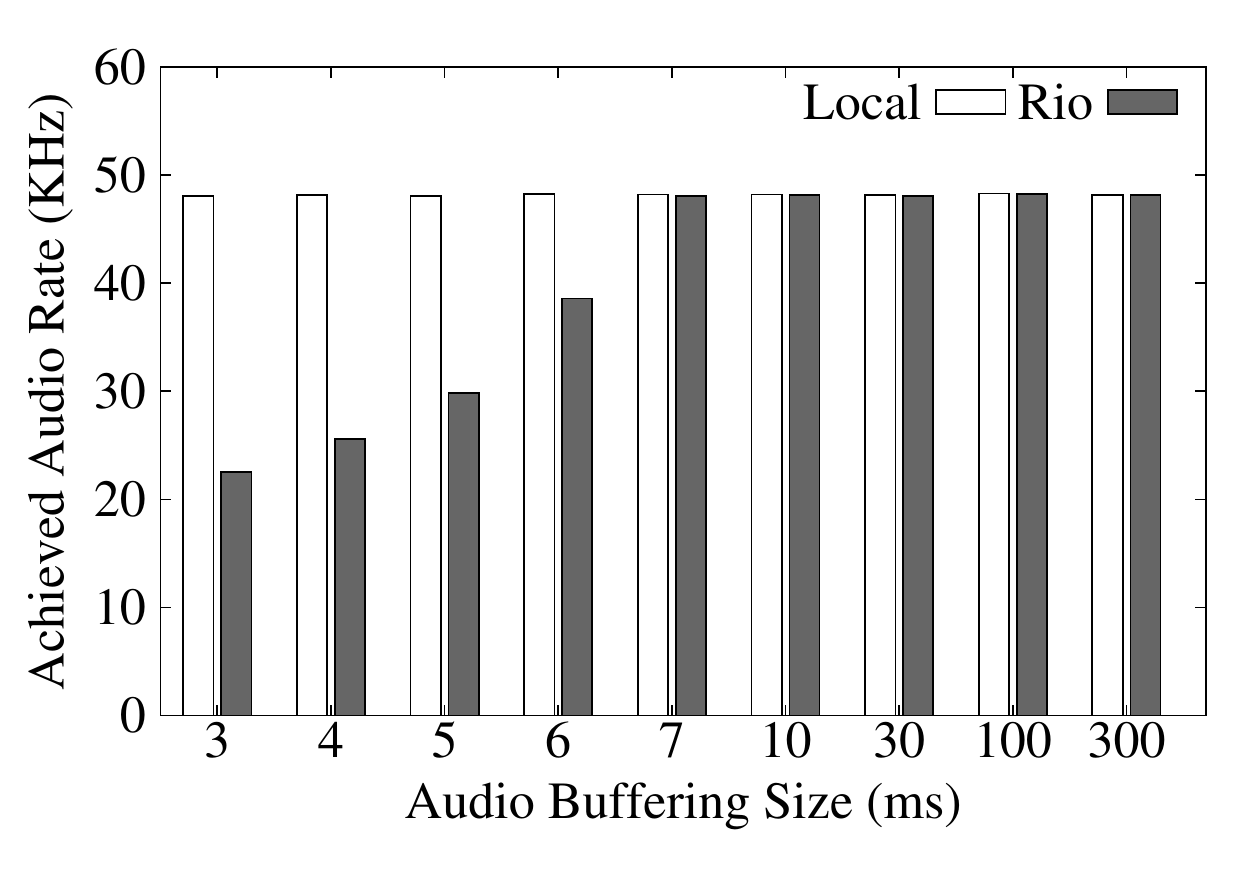}
}%
\hspace{0.5in}
\vspace{-2mm}
\subfigure[ ] { 
  \centering
  \includegraphics[width=0.9\columnwidth]{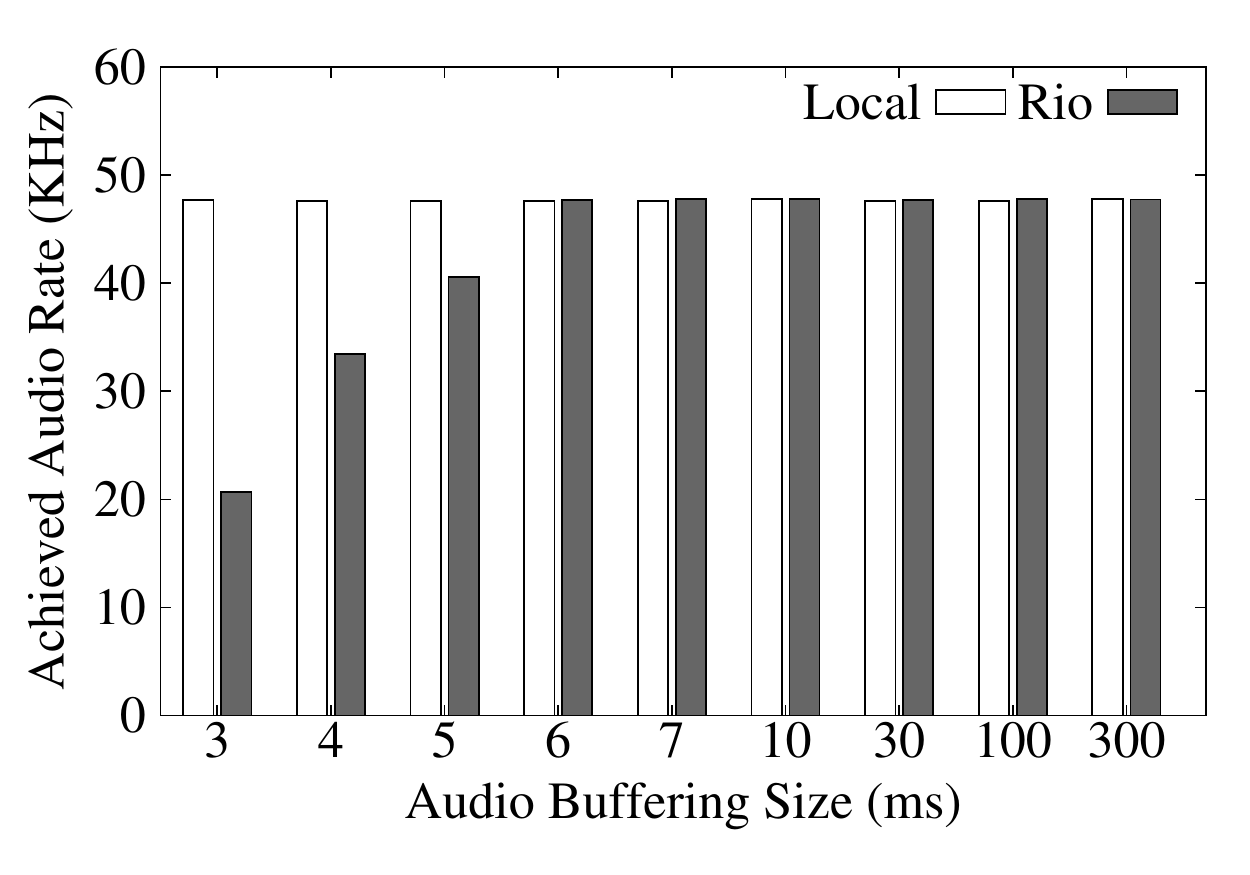}
}
\caption{Performance of speaker (a) and microphone (b). In both
figures, the X axis shows the buffering size in
the HAL. The larger the buffer size, the smoother the playback, but the larger the audio latency. The
Y axis shows the achieved audio rate.}
\label{fig:audio}
\end{figure*}

In this subsection, we measure the performance of different I/O classes
in Rio and compare them to local performance.

{\bf Audio devices:} We evaluate the performance of the speaker and
microphone by
measuring the audio sample rate at different buffering sizes, and hence,
different audio latency. 
Using larger buffering sizes reduces the interactions
with the driver but increases the audio latency. Audio latency is the average time it takes a sample to reach
the speaker from the process (and vice-versa for microphone), and is
directly determined by the buffering size used in the HAL. 


Figure~\ref{fig:audio} shows the achieved rate for different buffering
sizes (and hence different latencies) for the speaker and microphone when accessed
locally or remotely through Rio. 
We use a minimum of 3 ms for the buffering size as
it is the smallest size used for local speakers in Android in low
latency mode. The figure shows that
such a small buffering size degrades the audio rate in Rio. This is mainly
because the HAL issues one {\tt ioctl} for each 3 ms
 audio segment, but the {\tt ioctl} takes longer than 3 ms to
finish in Rio due to the network's high round-trip time. However, the
figure shows that Rio is able to achieve the desired 48 kHz audio rate
at a slightly larger buffering size of 6-7 ms. We believe that Rio
achieves acceptably low audio latency because Android uses
a buffering size of 308 ms for high latency audio mode for speaker, and
uses 22 ms
for microphone.

We also measure the performance of audio devices with Rio when mobile
systems are connected via the aforementioned high latency connection,
e.g., for making a phone call remotely at  work using a smartphone at
home.
Our measurements show that Rio  achieves the desired 48 kHz for the
microphone using buffering sizes as small as 85 ms. However, for the speaker, Rio can only
achieve a maximum sampling rate of 25 kHz using a 300 ms buffer (other buffer sizes performed poorly). 
While this is insufficient for stereo audio (which requires 48 kHz), it is adequate for mono audio.


\begin{figure*}[th]
\centering
\vspace{-2mm}
\subfigure[ ]{
  \centering
  \includegraphics[width=0.9\columnwidth]{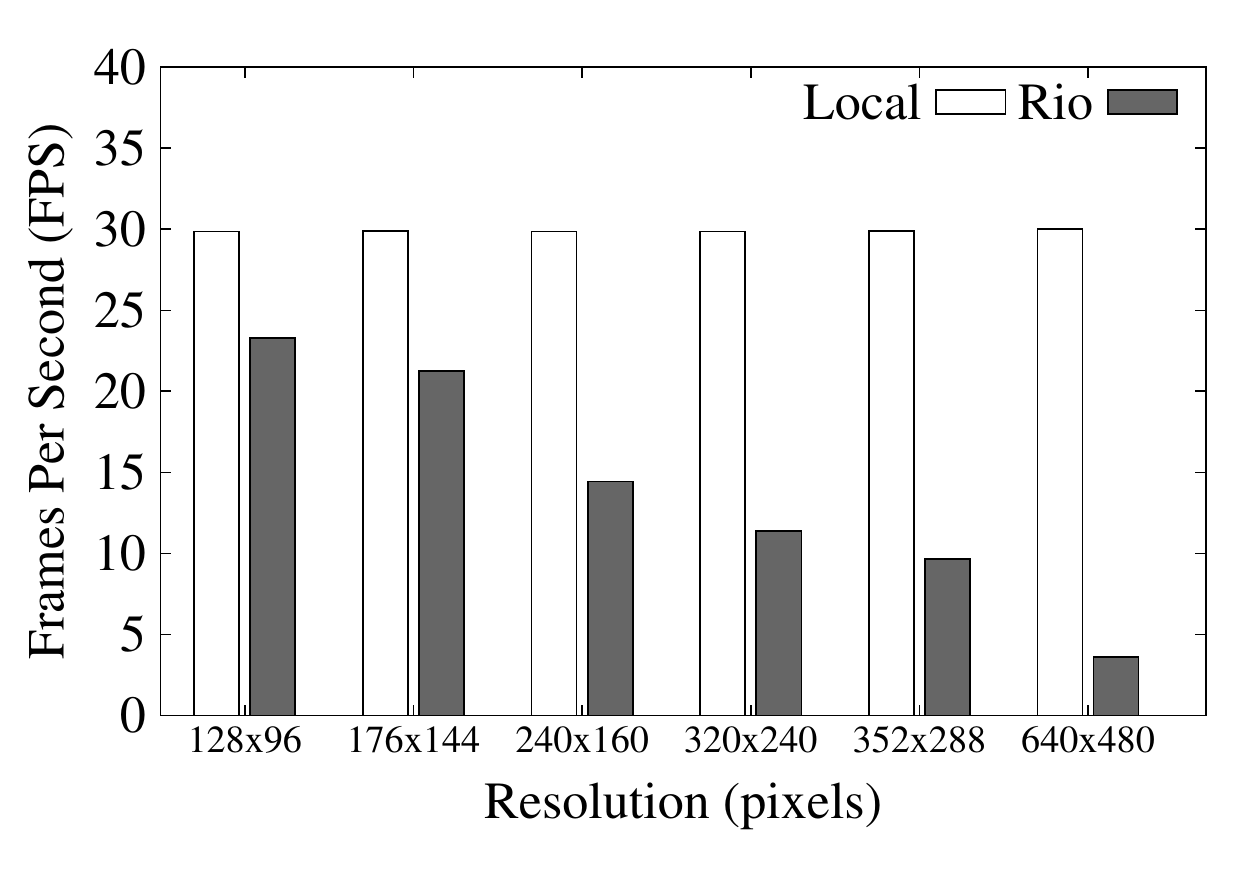}
}%
\hspace{0.5in}
\vspace{-2mm}
\subfigure[ ] { 
  \centering
  \includegraphics[width=0.9\columnwidth]{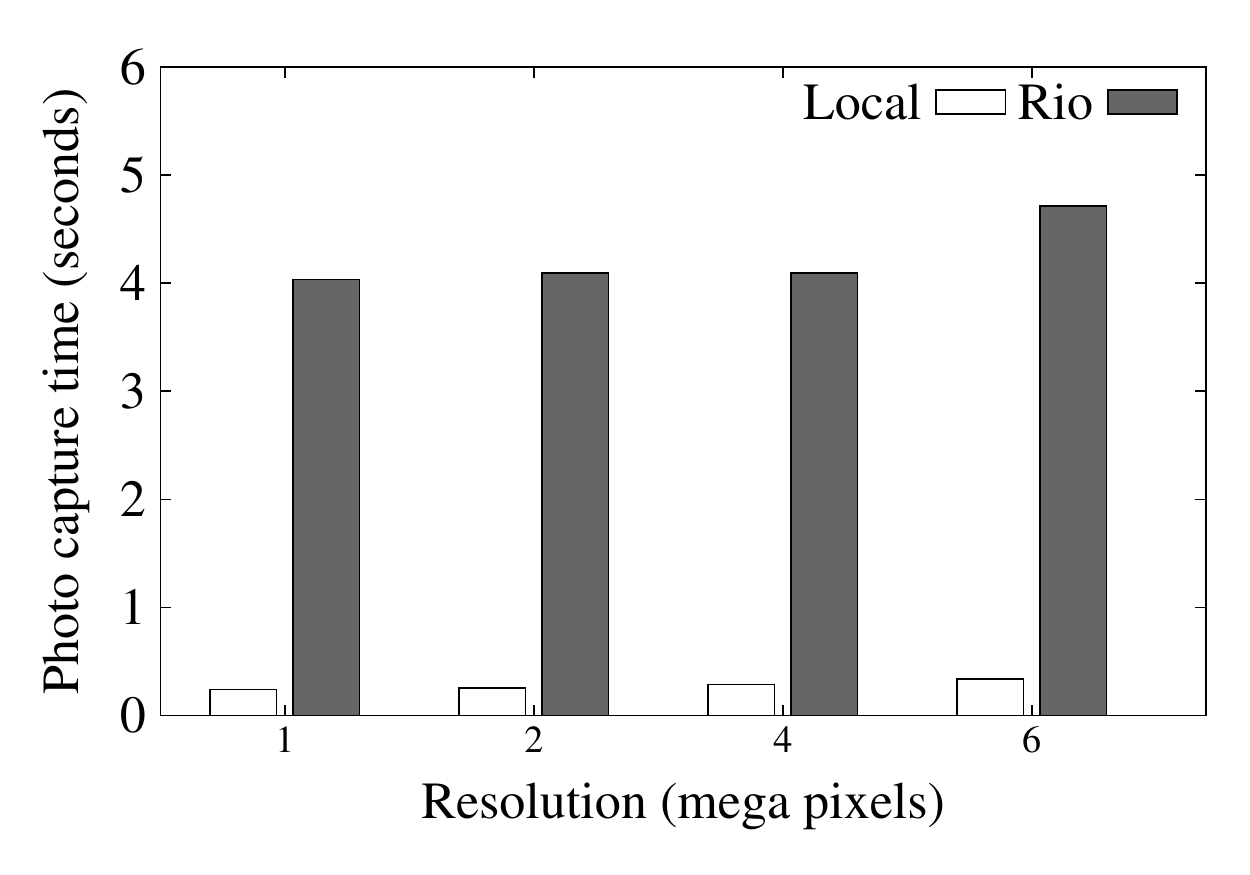}
}
\caption{Performance of a real-time streaming camera preview (a) and photo capture (b)
with a  14.3 Mbps wireless LAN connection between the client
and server. Future wireless standards with higher throughput will
improve  performance without requiring  changes to Rio.}
\label{fig:camera}
\end{figure*}

{\bf Camera:}
We measure the performance of both a real-time, streaming camera preview and the capture of a photo. 
In the first case, we measure the frame rate (in frames per
second) that the camera application can achieve, averaged over 1000
frames. 
We ignore the first 50 frames to avoid the effects of camera initialization
on performance. 

Figure~\ref{fig:camera}(a) shows that Rio can achieve acceptable performance (i.e., $>$15 FPS)
at low resolutions.
The performance at higher resolutions is bottlenecked by network throughput between the client and server.
Rio's efficient design spends most of its time transmitting frames rather than file operations.
However, streaming camera frames are uncompressed, requiring
612 KB of data {\em per frame} even for VGA (640$\times$480) resolution, necessitating 
about 72 Mbps of throughput to maintain 15 FPS.

We believe that the lower resolution camera preview supported by Rio 
is acceptable given that Rio supports capturing photos at maximum 
resolutions. Rio will support higher real-time
camera resolutions using future wireless standards; for example,
802.11n, 802.11ac, and a 802.11ad can achieve around 200 Mpbs, 600 Mbps, and
7 Gbps of throughput respectively~\cite{cisco_white_paper,
agilent_white_paper}. Such throughput capabilities  can support real-time camera
streaming in Rio at 15 FPS for resolutions of 1280$\times$720 and
1920$\times$1080, which are the highest
resolutions supported on Galaxy Nexus.

To evaluate photo capture, we measure the time from when the capture request is delivered to the
camera HAL from the application until the HAL notifies the application that the photo is ready. 
We do not include the focus time since it
is mainly dependent on the camera hardware and varies for different
scenes. Figure~\ref{fig:camera}(b) shows the capture time for local
and remote cameras using Rio. The figure shows the average over 10
captured photos for each resolution. It shows that Rio adds noticeable
latency to the capture time, mostly stemming from the time taken  to transfer the raw images from the server to the
client. However, the user only needs to point the camera at the targeted
scene very briefly (similar to when using a local camera), as the image will be
immediately captured in the server. 
It is important to note that the camera HAL in Galaxy Nexus
uses the same buffer size regardless of resolution, 
hence the capture time is essentially resolution-independent. 
The buffer size is 8 MB, which takes about 4.5 seconds to
transfer over our 14.3 Mbps connection. As with real-time camera streaming, 
future wireless standards will eliminate this overhead, providing latency on par with local camera capture.

{\bf Sensors:} To evaluate remote sensor performance, we measure the
average time it takes for the sensor HAL to obtain a new accelerometer
reading. Our measurements over 10000 samples show that the sensor HAL
obtains a new local reading in 65 ms on average and a new remote reading via Rio in 71 ms.
The sensor HAL obtains a new reading by issuing  a blocking {\tt poll} operation that waits in the
kernel until the data is ready, at which point the HAL issues a {\tt
read} file operation to read the new value. Rio causes overhead in this
situation because one and a half round trips 
are required since the blocking {\tt poll} returns until the {\tt read}
is completed.
Fortunately, this overhead is negligible in practice and does not impact the user experience.


{\bf Modem:} We measure the time it takes the dialer and messaging
applications to start a phone call and to send an SMS, respectively. 
We measure the time from when the user presses the ``dial'' or ``send SMS'' button
until the notification appears on the receiving phone. Our measurements
show that local and remote modems achieve similar performance, as the majority of time is
spent in carrier networks (from T-Mobile to AT\&T). For local and remote modems, The phone call takes about 7.8 and
7.9 seconds while SMS takes about 6.2 and 5.9 seconds, respectively. 

%
%
%

%
%
%


\section{Related Work}

The value of I/O sharing has been recognized by
others for both mobile and non-mobile systems. 
However, existing solutions have limitations: They do not support unmodified
applications, do not expose all the I/O device
functions to the client, or are I/O class-specific.

{\bf I/O sharing for mobile systems}:~~Existing I/O sharing solutions
for mobile systems all suffer from three fundamental limitations
described above.
For example, IP Webcam~\cite{IP_Webcam} turns
the camera on an Android system into an IP camera, which can then be
viewed from another mobile system through a custom viewer application.
The client system cannot configure most camera parameters, if any; all
configurations must be done manually on the server.  The client cannot
take photos either. Wi-Fi Speaker~\cite{Wi-Fi_Speaker} allows one to
play music on a smartphone from a PC. It does not allow the client to
configure the speaker parameters, e.g., the equalizer effects.
MightyText~\cite{MightyText} allows the user to send SMS and MMS
messages from a PC or a mobile system using the SIM card and modem in
another system. It does not support phone calls.

{\bf Screen sharing}:
Applications like Miracast~\cite{Miracast} allow
one system to send its screen for display on another.
Thin client solutions also display content received from
a server machine on a client. Examples are the X window
system~\cite{Scheifler1986}, THINC ~\cite{Baratto2004}, Microsoft Remote
Desktop~\cite{Cumberland99}, VNC~\cite{Richardson98}, Citrix
Metaframe~\cite{citrix}, and Sun Ray~\cite{Schmidt99}. None of these
solutions uses
the device file boundary and their choices of boundary are usually
graphics-specific or even application-specific. For example, X sets the
boundary between the application and X server. As a result, these
solutions will not support other classes of I/O devices

{\bf Other I/O sharing solutions}:~~
Remote file systems~\cite{NFS, Rifkin1986,
Leach 1997}, network USB devices~\cite{WebServicesOnDevices, Hari2011,
AnywhereUSB, USBOverIP, WirelessUSB},
Wireless Displays~\cite{WiDi}, and IP
cameras~\cite{Dropcam}
support I/O sharing as well. These solutions are also specific to one
I/O class, e.g., storage. Participatory sensing systems collect sensor data from registered mobile
systems~\cite{Das2010}. These systems use custom applications installed on mobile
systems, and therefore, are more limited than Rio, which supports
various I/O devices.

{\bf Computation offloading}:~~There is a large body of literature
regarding offloading computation from mobile
systems~\cite{flinn2012cyber}, e.g.,
Cyberforaging~\cite{balan2007mobisys}, MAUI~\cite{cuervo2010maui} and
COMET~\cite{gordon2012comet}. I/O sharing as is concerned in this work
invites a very different set of research challenges and a focus on
system support rather than programming support. Nevertheless, both
computation offloading and I/O sharing benefit from known techniques
from distributed systems. For example, both COMET and Rio employ DSM
albeit with very different designs.

\section{Concluding Remarks}
\label{sec:discussions_untrusted}

We presented Rio, an I/O sharing solution for mobile systems that adopts
a split-stack model at the device file boundary. We demonstrated that
Rio overcomes the limitations of existing solutions by supporting
unmodified
applications, exposing all I/O device functionality to clients, and
reducing development effort. We presented an implementation of Rio for
Android and showed that it achieves adequate performance for various
sharing scenarios and that it supports heterogeneous mobile systems.
We next offer some insights into the limitations of
the current design and implementation of Rio and our plans to overcome
some of them.

{\bf Supporting more classes of I/O devices}:~~Our current
implementation supports four classes of I/O devices. We plan to extend
it to support graphics, touchscreen, GPS, and FM radio, since they also
use the device file interface.
There are, however, two classes of I/O that Rio's design cannot support:
network and block devices. This is because these I/O devices do not use
the device file interface for communications between the process and the
driver. Network devices use sockets along with the kernel networking
stack and block devices use file systems.

{\bf Sharing I/O with untrusted systems}:~~
In this paper, we assumed that the systems sharing I/O through Rio trust
each other not to be malicious (\S\ref{sec:design_guarantees}).
Supporting untrusted systems creates new challenges for Rio, which fall into two
categories. \itemno{i} {\em Protecting the server.} As also discussed
in~\cite{AmiriSani2014}, device drivers are buggy and malicious
applications can abuse these bugs through the device file interface to
compromise the driver protection domain~\cite{BUG1, BUG2}. In Rio, this
means that a malicious process in the client can compromise the server,
e.g., through privilege escalation. In order to solve this problem, the
device driver and the device need to be sandboxed in a protection
domain in the server, using techniques similar to~\cite{AmiriSani2014}, 
Nooks~\cite{Swift2003}, and VirtuOS~\cite{Nikolaev2013}.
\itemno{ii} {\em Protecting the client.}  An untrusted
server can issue spurious copy memory operations to the client in order
to compromise the client.  The client stub can simply protect against
this threat by strictly checking the copy memory operations requested by
the server, similar to~\cite{AmiriSani2014}.
Note that the server can also snoop
the client's data that are shared with the I/O device,
e.g., the audio buffers.
Since the server is completely untrusted, we cannot
provide any isolation for the client's data. This is indeed an inherent
problem to any I/O sharing systems, and not specific to Rio.

{\bf Energy use by Rio}:~~Using an I/O device remotely via wireless
obviously incurs much more energy consumption than using a local one.
The device file boundary used by Rio is reasonably abstract that most of
the energy use by Rio comes from transporting the I/O data. Due to the
space limitation, we are not able to elaborate the energy optimizations
for Rio. Rather we note that most of the performance optimizations by
Rio, e.g., those described in \S\ref{sec:latency} lead to more efficient
use of the wireless and therefore to reduced energy consumption. We also
note that the paramount quest to reduce latency also rules out the use
of standard 802.11 power-saving mode with I/O sharing. On the other
hand, many known techniques that trade a little latency for much more
efficient use of wireless can benefit Rio, e.g., data
compression~\cite{barr2003energy} and $\mu$PM~\cite{liu2008mobisys}. 

{\bf Supporting iOS}:~~iOS also uses device files and hence can
be supported in Rio. Sharing I/O devices between iOS systems should
require similar engineering effort reported in this paper for sharing
I/O devices between Android systems.  However, sharing I/O devices
between iOS and Android systems require potentially non-trivial
engineering effort, mainly because these two systems have different I/O
stack components and hence different I/O device API.

\bibliographystyle{unsrt}
\bibliography{rio}

\begin{thebibliography}{10}

\bibitem{IP_Webcam}
{Android IP Webcam application}.
\newblock
  \url{https://play.google.com/store/apps/details?id=com.pas.webcam&hl=en}.

\bibitem{Wi-Fi_Speaker}
{Android Wi-Fi Speaker application}.
\newblock
  \url{https://play.google.com/store/apps/details?id=pixelface.android.audio&h%
l=en}.

\bibitem{MightyText}
{MightyText application}.
\newblock \url{http://mightytext.net}.

\bibitem{Miracast}
{Miracast}.
\newblock
  \url{https://www.wi-fi.org/sites/default/files/uploads/wp_Miracast_Industry_%
20120919.pdf}.

\bibitem{AmiriSani2014}
A.~Amiri~Sani, K.~Boos, S.~Qin, and L.~Zhong.
\newblock {I/O Paravirtualization at the Device File Boundary}.
\newblock {\em to Appear in Proc. ACM ASPLOS}, 2014.

\bibitem{rio_page}
{Rio Project Homepage (including a video demo)}.
\newblock \url{http://www.ruf.rice.edu/~mobile/rio.html}.

\bibitem{li88}
Kai Li.
\newblock Ivy: A shared virtual memory system for parallel computing.
\newblock In {\em Proc. 1988 Int. Conf. Parallel Processing}, volume~2, pages
  94--101, 1988.

\bibitem{delp88}
Gary~Scott Delp.
\newblock The architecture and implementation of memnet: a high--speed
  shared-memory computer communication network.
\newblock 1988.

\bibitem{lenoski90}
Daniel Lenoski, James Laudon, Kourosh Gharachorloo, Anoop Gupta, and John
  Hennessy.
\newblock {\em The directory-based cache coherence protocol for the DASH
  multiprocessor}, volume~18.
\newblock ACM, 1990.

\bibitem{Carter91}
John~B. Carter, John~K. Bennett, and Willy Zwaenepoel.
\newblock {Implementation and Performance of Munin}.
\newblock In {\em Proc. ACM SOSP}.

\bibitem{zhou92}
Songnian Zhou, Michael Stumm, Kai Li, and David Wortman.
\newblock Heterogeneous distributed shared memory.
\newblock {\em Parallel and Distributed Systems, IEEE Transactions on},
  3(5):540--554, 1992.

\bibitem{schoinas94}
Ioannis Schoinas, Babak Falsafi, Alvin~R Lebeck, Steven~K Reinhardt, James~R
  Larus, and David~A Wood.
\newblock {\em Fine-grain access control for distributed shared memory},
  volume~29.
\newblock ACM, 1994.

\bibitem{ARM_TRM}
ARM.
\newblock {Architecture Reference Manual, ARMv7-A and ARMv7-R edition}.
\newblock {\em ARM DDI}, 0406A, 2007.

\bibitem{Woodings2006}
Ryan Woodings and Manoj Pandey.
\newblock {WirelessUSB: a Low Power, Low Latency and Interference Immune
  Wireless Standard}.
\newblock In {\em Proc. IEEE Wireless Communications and Networking Conference
  (WCNC)}, 2006.

\bibitem{ksocket}
{Linux ksocket}.
\newblock \url{http://ksocket.sourceforge.net/}.

\bibitem{OMAP_TRM}
Texas Instruments.
\newblock {Architecture Reference Manual, OMAP4430 Multimedia Device Silicon
  Revision 2.x}.
\newblock SWPU231N, 2010.

\bibitem{ION}
{Android ION Memory Allocator}.
\newblock \url{http://lwn.net/Articles/480055/}.

\bibitem{self_photos}
{Applications for taking self photos}.
\newblock
  \url{http://giveawaytuesdays.wonderhowto.com/inspiration/10-iphone-and-andro%
id-apps-for-taking-self-portraits-0129658/}.

\bibitem{SIM_lock}
\url{http://www.theverge.com/2011/06/09/google-voice-skype-imessage-and-the-de%
ath-of-the-phone-number/}.

\bibitem{Raskar2006}
R.~Raskar, J.~Tumblin, A.~Mohan, A.~Agrawal, and Y.~Li.
\newblock {Computational Photography}.
\newblock In {\em Proc. STAR Eurographics}, 2006.

\bibitem{Yuan2007}
Lu~Yuan, Jian Sun, Long Quan, and Heung-Yeung Shum.
\newblock {Image Deblurring with Blurred/Noisy Image Pairs}.
\newblock {\em ACM Transactions on Graphics (TOG)}, 26(3):1, 2007.

\bibitem{Wilburn2005}
Bennett Wilburn, Neel Joshi, Vaibhav Vaish, Eino-Ville Talvala, Emilio Antunez,
  Adam Barth, Andrew Adams, Mark Horowitz, and Marc Levoy.
\newblock {High Performance Imaging Using Large Camera Arrays}.
\newblock {\em ACM Transactions on Graphics (TOG)}, 24(3):765--776, 2005.

\bibitem{Park2003}
Sung~Cheol Park, Min~Kyu Park, and Moon~Gi Kang.
\newblock {Super-Resolution Image Reconstruction: a Technical Overview}.
\newblock {\em IEEE Signal Processing Magazine}, 20(3):21--36, 2003.

\bibitem{cisco_white_paper}
{802.11ac: The Fifth Generation of Wi-Fi}.
\newblock In {\em Cisco White Paper}, 2012.

\bibitem{agilent_white_paper}
{Wireless LAN at 60 GHz - IEEE 802.11ad Explained}.
\newblock In {\em Agilent White Paper}.

\bibitem{Scheifler1986}
R.W. Scheifler and J.~Gettys.
\newblock The {X} window system.
\newblock {\em ACM Transactions on Graphics}, 5(2), 1986.

\bibitem{Baratto2004}
R.A. Baratto, L.~Kim, and J.~Nieh.
\newblock Thinc: A remote display architecture for thin-client computing.
\newblock In {\em Proc. ACM SOSP}, 2004.

\bibitem{Cumberland99}
B.C. Cumberland, G.~Carius, and A.~Muir.
\newblock {\em Microsoft Windows NT Server 4.0, Terminal Server Edition:
  Technical Reference}.
\newblock Microsoft Press, 1999.

\bibitem{Richardson98}
T.~Richardson, Q.~Stafford-Fraser, K.R. Wood, and A.~Hopper.
\newblock Virtual network computing.
\newblock {\em IEEE Internet Computing}, 1998.

\bibitem{citrix}
{Citrix Metaframe}.
\newblock \url{http://www.citrix.com}.

\bibitem{Schmidt99}
Brian~K. Schmidt, Monica~S. Lam, and J.~Duane Northcutt.
\newblock The interactive performance of slim: A stateless, thin-client
  architecture.
\newblock In {\em Proc. ACM SOSP}, 1999.

\bibitem{NFS}
{Network File System}.
\newblock \url{http://etherpad.tools.ietf.org/html/rfc3530}.

\bibitem{Rifkin1986}
A.P. Rifkin, M.P. Forbes, R.L. Hamilton, M.~Sabrio, S.~Shah, and K.~Yueh.
\newblock Rfs architectural overview.
\newblock In {\em Proc. USENIX Conference}, 1986.

\bibitem{Leach1997}
P.J. Leach and D.~Naik.
\newblock A common internet file system (cifs/1.0) protocol.
\newblock {\em Draft, Network Working Group, IETF}, 1997.

\bibitem{WebServicesOnDevices}
Web services on devices: Devices that are controlled on the network.
\newblock
  \url{http://msdn.microsoft.com/en-us/library/windows/desktop/aa826001(v=vs.8%
5).aspx}.

\bibitem{Hari2011}
A.~Hari, M.~Jaitly, Y.J. Chang, and A.~Francini.
\newblock The switch army smartphone: Cloud-based delivery of {USB} services.
\newblock In {\em Proc. ACM MobiHeld}, 2011.

\bibitem{AnywhereUSB}
{Digi International: AnywhereUSB}.
\newblock \url{http://www.digi.com/products/usb/anywhereusb.jsp}.

\bibitem{USBOverIP}
{USB Over IP}.
\newblock \url{http://usbip.sourceforge.net/}.

\bibitem{WirelessUSB}
{Wireless USB}.
\newblock \url{http://www.usb.org/wusb/home/}.

\bibitem{WiDi}
{Intel WiDi}.
\newblock
  \url{http://www.intel.com/content/www/us/en/architecture-and-technology/inte%
l-wireless-display.html}.

\bibitem{Dropcam}
{Dropcam}.
\newblock \url{https://www.dropcam.com/}.

\bibitem{Das2010}
T.~Das, P.~Mohan, V.~Padmanabhan, R.~Ramjee, and A.~Sharma.
\newblock {PRISM}: Platform for remote sensing using smartphones.
\newblock In {\em Proc. ACM MobiSys}, 2010.

\bibitem{flinn2012cyber}
Jason Flinn.
\newblock Cyber foraging: Bridging mobile and cloud computing.
\newblock {\em Synthesis Lectures on Mobile and Pervasive Computing}, 2012.

\bibitem{balan2007mobisys}
Rajesh~Krishna Balan, Darren Gergle, Mahadev Satyanarayanan, and James
  Herbsleb.
\newblock Simplifying cyber foraging for mobile devices.
\newblock In {\em Proc. ACM MobiSys}, 2007.

\bibitem{cuervo2010maui}
Eduardo Cuervo, Aruna Balasubramanian, Dae-ki Cho, Alec Wolman, Stefan Saroiu,
  Ranveer Chandra, and Paramvir Bahl.
\newblock Maui: making smartphones last longer with code offload.
\newblock In {\em Proc. ACM MobiSys}, 2010.

\bibitem{gordon2012comet}
Mark~S Gordon, D~Anoushe Jamshidi, Scott Mahlke, Z~Morley Mao, and Xu~Chen.
\newblock Comet: code offload by migrating execution transparently.
\newblock In {\em Proc. OSDI}, 2012.

\bibitem{BUG1}
{Privilege escalation using NVIDIA GPU driver bug}.
\newblock \url{http://www.securelist.com/en/advisories/50085}.

\bibitem{BUG2}
{Privilege escalation using DRM/Radeon GPU driver bug}.
\newblock \url{https://lkml.org/lkml/2010/1/18/106}.

\bibitem{Swift2003}
Michael~M Swift, Brian~N Bershad, and Henry~M Levy.
\newblock Improving the reliability of commodity operating systems.
\newblock In {\em Proc. ACM SOSP}, 2003.

\bibitem{Nikolaev2013}
Ruslan Nikolaev and Godmar Back.
\newblock {VirtuOS: An Operating System with Kernel Virtualization}.
\newblock In {\em Proc. ACM SOSP}, 2013.

\bibitem{barr2003energy}
Kenneth~C Barr and Krste Asanovi{\'c}.
\newblock Energy-aware lossless data compression.
\newblock In {\em Proc. ACM MobiSys}, 2003.

\bibitem{liu2008mobisys}
Jiayang Liu and Lin Zhong.
\newblock Micro power management of active 802.11 interfaces.
\newblock In {\em Proc. ACM MobiSys}, 2008.

\end{thebibliography}


\balancecolumns
\end{document}